\documentclass[aps,prl,superscriptaddress,twocolumn,showpacs,nofootinbib,titlepage=false]{revtex4}
\usepackage{graphicx}
\usepackage{amsmath}
\usepackage{amssymb}
\usepackage{subfigure}
\usepackage{natbib}

\def\ket#1{{\left| #1 \right\rangle}}

\begin{document}

\title{Multi-mode Entanglement is Detrimental to Lossy Optical Quantum Metrology}
\author{P.A. Knott}
	\email{phy5pak@leeds.ac.uk}
	\affiliation{School of Physics and Astronomy, University of Leeds, Leeds LS2 9JT, United Kingdom}
	\affiliation{NTT Basic Research Laboratories, NTT Corporation, 3-1 Morinosato-Wakamiya, Atsugi, Kanagawa 243-0198, Japan}
\author{T.J. Proctor}
	\affiliation{School of Physics and Astronomy, University of Leeds, Leeds LS2 9JT, United Kingdom}
\author{Kae Nemoto}
	\affiliation{National Institute of Informatics, 2-1-2 Hitotsubashi, Chiyoda-ku, Tokyo 101-8430, Japan}
\author{J.A. Dunningham}
	\affiliation{Department of Physics and Astronomy, University of Sussex, Brighton BN1 9QH, United Kingdom}
\author{W.J. Munro}
	\affiliation{NTT Basic Research Laboratories, NTT Corporation, 3-1 Morinosato-Wakamiya, Atsugi, Kanagawa 243-0198, Japan}
	\affiliation{National Institute of Informatics, 2-1-2 Hitotsubashi, Chiyoda-ku, Tokyo 101-8430, Japan}
\pacs{42.50.St,42.50.Dv,03.65.Ud,03.65.Ta}

\date{\today}

\begin{abstract}

In optical interferometry multi-mode entanglement is often assumed to be the driving force behind quantum enhanced measurements. Recent work has shown this assumption to be false: single mode quantum states perform just as well as their multi-mode entangled counterparts. We go beyond this to show that when photon losses occur - an inevitability in any realistic system - multi-mode entanglement is actually detrimental to obtaining quantum enhanced measurements. We specifically apply this idea to a superposition of coherent states, demonstrating that these states show a robustness to loss that allows them to significantly outperform their competitors in realistic systems. A practically viable measurement scheme is then presented that allows measurements close to the theoretical bound, even with loss. These results promote a new way of approaching optical quantum metrology using single-mode states that we expect to have great implications for the future.

\end{abstract}
\maketitle



In recent years quantum metrology has begun to fulfil its potential as an important practical method of enhancing precision measurements \cite{caves1982quantum,grote2013first,taylor2013biological,giovannetti2004quantum,nagata2007beating}. It has a wide range of applications, from time and frequency measurements \cite{flowers2004route,PhysRevLett793865} to lithography \cite{boto2000quantum,d2004two}, and is already being used to surpass the classical limit in gravitational wave detectors \cite{abadie2011gravitational,aasi2013enhanced,schnabel2010quantum}. The precision gains offered by quantum metrology are often attributed to entanglement \cite{lee2002quantum,afek2010high,giovannetti2006quantum,cappellaro2005entanglement,riedel2010atom} and, specifically in the optical case, entanglement between two modes in an interferometer \cite{jin2013sequential,kok2002creation}. However, more recently it has been shown that entanglement is not required between the probe and reference systems for Heisenberg limited measurements of a linear phase shift  \cite{munro2001weak,ralph2002coherent,tilma2010entanglement}, and furthermore it has been argued that the important resource for enhancing precision is actually the coherence in the eigenbasis of the phase shift Hamiltonian \cite{girolami2013characterizing}. We go beyond this to demonstrate that, in some well known scenarios, single-mode superposition states have a significantly better robustness to loss than their multi-mode entangled counterparts, which allows them to achieve greatly enhanced precision measurements. We introduce a new state, the unbalanced cat state, that can outperform the alternatives and can be created and measured with present day or near future technology, to a precision close to its theoretical bound. We begin by examining the ideal case where no photon losses are present.

\emph{No loss: superposition states are sufficient -} The quantum fisher information (QFI), which quantifies a state's ability to measure a phase $\phi$, for a general density matrix $\rho$ is given by \citep{braunstein1994statistical,braunstein1996generalized,joo2011quantum}:
\begin{eqnarray}
F_Q = \sum_{i,j} \frac{2}{\lambda_i+\lambda_j}\left| \langle \lambda_i | \partial \rho(\phi) / \partial \phi| \lambda_j \rangle\right|^2,
\label{QFI}
\end{eqnarray}
where $\lambda_i$ are the eigenvalues and $|\lambda_i \rangle$ a corresponding set of orthonormal eigenvectors of $\rho$. For a pure state $|\Psi\rangle$ the QFI is \cite{demkowicz2009quantum}: $F_Q = 4 \left[  \langle \Psi ' | \Psi ' \rangle - | \langle \Psi ' | \Psi \rangle |^2 \right]$, where $| \Psi ' \rangle = \frac{\partial}{\partial \phi} | \Psi  \rangle$. The fundamental limit to the precision with which the state $\rho$ can measure a phase $\phi$ is then given by the quantum Cram\'er-Rao bound (CRB) \citep{braunstein1994statistical,braunstein1996generalized}:
\begin{eqnarray}
\delta \phi \ge \frac{1}{\sqrt{m F_Q}}, \label{CRB}
\end{eqnarray}
where $m$ is the number of times that the measurement is independently repeated. From this it is straightforward to show that a linear phase measurement involving $N$ independent particles gives a precision at the shot noise limit (SNL), given by $\delta\phi = 1/\sqrt{N}$ \citep{footnote1}.

A well studied state for quantum-enhanced metrology is the NOON state \citep{lee2002quantum,afek2010high,israel2014supersensitive} given by $|\Psi_{NOON}\rangle = {1 \over \sqrt{2}} ( |N,0\rangle_{1,2} + |0,N\rangle_{1,2} )$, where the subscripts refer to two different modes in a Mach-Zehnder interferometer. This state is maximally entangled and, using equation \ref{CRB}, it can be shown that the NOON state can measure a phase with a quantum-enhanced precision of $\delta\phi_{NOON}=1/N$, the Heisenberg limit. However, the Heisenberg limit is attainable without the multi-mode entanglement exhibited by the NOON state, simply by utilizing an analogous single mode superposition state $|\Psi_{NO}\rangle= {1 \over \sqrt{2}} ( |N\rangle + |0\rangle )$ \citep{tilma2010entanglement}, which we refer to as the NO state.

An alternative state that has been shown to be useful for quantum metrology is the entangled coherent state (ECS) $\ket{\Psi_{ECS}} =\mathcal{N}_e ( |\alpha,0\rangle_{1,2} + |0,\alpha \rangle_{1,2} )$ \citep{sanders1992entangled,gerry1997generation,gerry2002nonlinear,gerry2007nonlocal,gerry2009maximally,gerry2010heisenberg,joo2011quantum,joo2012quantum,sanders2012review} where $\mathcal{N}_e=1/\sqrt{2+2e^{-\alpha^2}}$ and $\alpha$ characterizes the coherent state (we take $\alpha$ to be real throughout without loss of generality). The QFI for this state is given by:
\begin{eqnarray}
\label{eq:QFI_ECS_Cat}
F_Q=4\alpha^2 \mathcal{N}_e^2 (1+\alpha^2-\alpha^2\mathcal{N}_e^2),
\end{eqnarray}
which approximately scales as $F_Q \propto \alpha^4$. However, a very similar QFI can be obtained without the entanglement by utilizing the single mode analogue of the ECS - a balanced cat state (equally weighted superposition of two coherent states) - given by $\ket{\Psi_{cat}}=\mathcal{N}_c ( |\alpha\rangle + |0\rangle ) $ where $\mathcal{N}_c=1/\sqrt{2+2e^{-\alpha^2/2}}$. Its QFI is also given by Eq. (\ref{eq:QFI_ECS_Cat}) but with $\mathcal{N}_e$ replaced with $\mathcal{N}_c$.

\begin{figure}[t]
\centering
\includegraphics[scale=1.2]{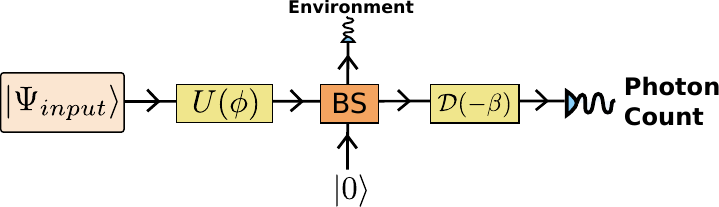}
\caption{(color online) The input state $\ket{\Psi_{input}}$ first undergoes a phase shift $U(\phi)=e^{i\hat{a}^{\dagger}\hat{a}\phi}$. We model loss by the addition of a beam splitter with a vacuum port, and then trace over the environmental mode. To read out the phase we apply the displacement operator $\mathcal{D}(-\beta)$, with coherent state amplitude $-\beta$, and then count the number of photons in the state.}
\label{fig:scheme}
\end{figure}

\begin{figure}[t]
\centering
\includegraphics[scale=0.47]{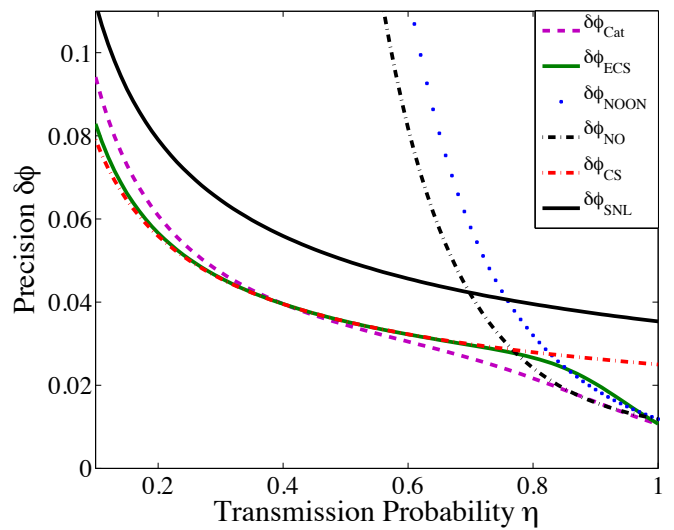}
\caption{(color online) Multi-mode entanglement reduces phase precision. We show the CRB, calculated from the QFI, for the: cat state $\delta\phi_{\text{cat}}$; ECS $\delta\phi_{\text{ECS}}$; NOON state $\delta\phi_{\text{NOON}}$; NO state $\delta\phi_{\text{NO}}$; coherent state $\delta\phi_{\text{CS}}$ and the SNL $\delta\phi_{\text{SNL}}$. Here $\alpha=3$, and for fair comparison the NOON and NO states have $N$ such that the number of photons per state through the phase shift $\bar{n}_{\phi}$ is equal for each state. Therefore $\bar{n}_{\phi}(\text{NOON})=\bar{n}_{\phi}(\text{NO})=N/2$ is equal to $\bar{n}_{\phi}(\text{ECS})=\bar{n}_{\phi}(\text{cat})=\mathcal{N}^2\alpha^2$. We repeat each state $m$ times so that the total number of photons sent through the phase shift is $R_{\phi}=m\bar{n}_{\phi}=400$ (this is the same throughout our results).}
\label{fig:al3_part_graph1_b}
\end{figure}

\emph{Entanglement is detrimental with loss -} We now show that in the presence of photon losses the single-mode states have a significant improvement in phase sensitivity over their multi-mode analogues. We model loss by the addition of a beam splitter after the phase shift \citep{gkortsilas2012measuring,joo2011quantum,demkowicz2009quantum}, as shown in Fig.~\ref{fig:scheme}, which has a probability of transmission $\eta$ (and therefore a fraction $\mu=1-\eta$ of photons are lost). After tracing over the environment we have a mixed state $\rho$, and from this density matrix the QFI can be determined. The CRB as a function of loss has been calculated for the NOON state \citep{demkowicz2009quantum} and the ECS \citep{zhang2013quantum}, and we have calculated this for the NO and cat states. The results, in Fig.~\ref{fig:al3_part_graph1_b}, show that with loss the NO state ($\delta\phi_{\text{NO}}$, black dashed-dotted line) can measure a phase to a higher precision than a NOON state ($\delta\phi_{\text{NOON}}$, blue dots). However, it is not clear how to create a NO state in a physically viable fashion. We note that the NOON state has the same QFI as the NO state if there is only loss in the phase shift arm, highlighting the similarity between single mode metrology and a two mode scheme with loss only at the phase shift.

We can see from Fig.~\ref{fig:al3_part_graph1_b} that in the range of reasonable experimental transmission rates, $0.5 \le \eta \le 1$, (for example $\eta=0.62$ in \citep{demkowicz2013fundamental}) the precision obtained by the cat state ($\delta\phi_{\text{cat}}$, purple dashed line) is significantly better than the ECS ($\delta\phi_{\text{ECS}}$, green solid line). In this region the multi-mode entanglement in the ECS leads to a more fragile state and a worse precision. Despite this, we can see that for higher loss rates the ECS performs better than the cat state. We now show that the single mode states can be modified to overcome this issue.

\emph{The unbalanced cat state -} We now introduce a single mode state that generalizes the cat state and displays an improvement in phase sensitivity over the ECS (and the other alternatives) for all values of loss. We will refer to the state as the unbalanced cat state (UCS) and it is given by:
\begin{equation}
|\Psi_{UCS}\rangle=\mathcal{N}_{u}(|\alpha(a)\rangle+a|0\rangle ) \label{UCSstate}
\end{equation}
where $0 \leq a \leq 1$, $\mathcal{N}_{u} = 1/\sqrt{1+a^2+2ae^{-\alpha(a)^2/2}}$, and $\alpha(a)$ is the solution to: $\alpha^2(a) = \bar{n}_{\phi} / \mathcal{N}^2_{u}(\alpha(a))$, where $\bar{n}_{\phi}$ is the number of photons passing through the phase shift per state. $\alpha(a)$ is defined in such a way as to keep the average number of photons through the phase shift independent of $a$, and it can be expressed in terms of the Lambert W-function. We note that taking $a=1$ in Eq.~(\ref{UCSstate}) gives a balanced cat state of magnitude $\alpha_{bal}=\alpha(a=1)$, and $a=0$ gives a coherent state. One of the advantages of this state is that the `quantumness' of the state can be altered by varying the parameter $a$. Loss collapses the quantum superposition, and so when there is high loss we can reduce $a$ so that the state behaves more like a coherent state $|\alpha\rangle$, and with low loss we can set $a \sim 1$ so that we have an equal superposition state. 

\begin{figure}[t]
\centering
\includegraphics[scale=0.47]{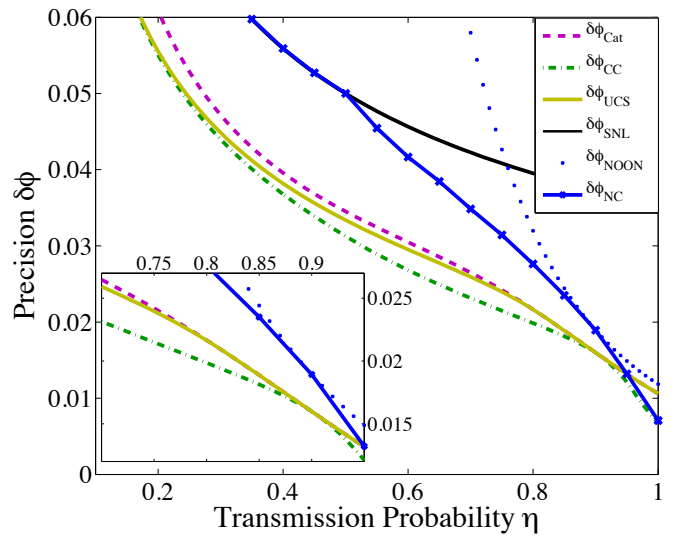}
\caption{(color online) We see the large improvements gained by our single mode states. We show the CRB for the: cat state $\delta\phi_{\text{cat}}$; UCS chopping $\delta\phi_{\text{CC}}$; unbalanced cat state $\delta\phi_{\text{UCS}}$; SNL $\delta\phi_{\text{SNL}}$; NOON state $\delta\phi_{\text{NOON}}$ and the NOON chopping strategy $\delta\phi_{\text{NC}}$. Here $\alpha_{bal}=3$, and for the chopped states we limit the cat state to $\alpha_{bal} \le 5$ (and equivalently limit the NOON state).}
\label{fig:al3_part_graph2_b}
\end{figure}

The reduced density matrix $\rho_{_{UCS}}$ for the UCS after the phase shift and loss is given by:
\begin{align}
\rho_{_{UCS}} &= \mathcal{N}_{u}^2 \left[ |\alpha_{\eta}(a)e^{i\phi}\rangle\langle \alpha_{\eta}(a)e^{i\phi}| + a^2|0\rangle\langle0| \vphantom{e^{-\alpha^{2}_{\mu}(a)/2}}\right. \\
  &\quad \left. + a e^{-\alpha^{2}_{\mu}(a)/2} \left( |\alpha_{\eta}(a)e^{i\phi} \rangle\langle 0 | + |0\rangle\langle \alpha_{\eta}(a)e^{i\phi}| \right) \right], \notag
  \label{rho}
\end{align}
where $\alpha_{\eta}(a)=\alpha(a) \sqrt{\eta}$ and $\alpha_{\mu}(a)=\alpha(a) \sqrt{\mu}$. Using a similar method to \citep{zhang2013quantum} we can represent and then diagonalize $\rho_{_{UCS}}$ in the orthogonal cat state basis $| \Psi_{\pm} \rangle = \mathcal{N}_{\pm} (| \alpha_{\eta}(a) e^{i\phi} \rangle \pm |0 \rangle )$  to find the two nonzero eigenvalues and the corresponding eigenvectors. Using Eq.~(\ref{QFI}) and Eq.~(\ref{CRB}) we then calculate the QFI and the CRB, and optimize this over the range of possible choices of $a$ for each value of loss.
\newline
\indent
We see in Fig.~\ref{fig:al3_part_graph2_b} that the CRB for the UCS ($\delta\phi_{\text{UCS}}$, yellow solid line) improves upon the cat state ($\delta\phi_{\text{cat}}$, purple dashed line). Although this improvement in the CRB is marginal, we will show that with a simple and practical measurement scheme the UCS, unlike the balanced cat state, can be utilized for phase measurements close to the CRB. We note that both the cat and the UCS show large precision improvements over the SNL.
\newline
\indent
We can obtain a better precision still by using a `chopping strategy', introduced in the case of NOON states in \citep{dorner2009optimal}, in which different sized UCSs (i.e. different $\bar{n}_{\phi}$) are used for different loss rates. We fix the total number of photons allowed through the phase shift, $R_{\phi}$, and therefore the number of times $m$ that a state is sent through the phase shift is inversely proportional to its average photon number $\bar{n}_{\phi}$. The green dashed-dotted line ($\delta\phi_{\text{CC}}$) in Fig.~\ref{fig:al3_part_graph2_b} shows a UCS optimized over $\bar{n}_{\phi}$ and the unbalancing parameter $a$ for each loss rate. The chopping strategy utilises the fact that larger states obtain higher precsion for low loss, whereas for high loss smaller states are more robust and are therefore preferable. For this reason, when there is no loss it is advantageous to take the largest possible state, which we limit here to having magnitiude $\alpha_{bal} = 5$, as larger states than this are physically unrealistic. We see that this chopping strategy applied to the unbalanced cat displays further improvements over all the alternatives, including a vast improvement over the NOON chopping strategy ($\delta \phi_{\text{NC}}$) and the SNL ($\delta \phi_{\text{SNL}}$).

By looking at the theoretical limits on the precision (given by the CRB) for various single-mode states, it is clear that these states have huge potential for making quantum-enhanced measurements. Despite this, it is not always clear how to make measurements that saturate this limit, and it is this issue that we turn to next.

\emph{A measurement scheme for the UCS with loss -} 
We will now describe a simple and practical scheme, shown in Fig.~\ref{fig:scheme}, for measuring a phase using a UCS, in the presence of loss, that comes close to the theoretical precision limit given by the CRB. The initial resource required is a UCS. There are many examples of cat state generation techniques, such as that given in \citep{brune1996observing}. In this scheme a Rydberg atom in a cavity in the state $|g \rangle + |e \rangle$ is coupled to a coherent state via the Jaynes-Cummings Hamiltonian \citep{jaynes1963comparison}. The atom-cavity system evolves as $|\alpha' \rangle (|g \rangle + |e \rangle) \rightarrow |\alpha' \rangle |g \rangle + |\alpha' e^{i\phi} \rangle |e \rangle$ and, after a transformation and measurement of the Rydberg atom and taking $\phi=\pi$, the resultant state of the field is an even cat state. Alternative schemes are numerous \citep{lund2004conditional,bartley2012multiphoton,gerrits2010generation,leghtas2013deterministic}, and cat states have been created with $\alpha'=1.76$ and fidelity $0.59$ in the lab \citep{gerrits2010generation}. 
Some schemes create states of the form $|\Psi_{cat}\rangle=\mathcal{N}_c ( |\alpha\rangle + |0\rangle ) $ directly \citep{leghtas2013deterministic}, but if the output state is $\mathcal{N}_{c'}(|\alpha'\rangle +|-\alpha'\rangle)$ the application of a displacement operator \citep{paris1996displacement} will create the state $|\Psi_{cat}\rangle$. The UCS can be created by simple adaptations of these methods for cat state preparation, for example preparing the Rydberg atom in the unbalanced state $\mathcal{N}_R(|g \rangle + a|e \rangle)$ will give the output state $\ket{\Psi_{UCS}}$. 
\newline
\indent
The first step in the phase detection scheme is the application of the linear phase shift to the UCS giving $|\Psi_{UCS}(\phi)\rangle=\mathcal{N}_c(|\alpha(a) e^{i\phi}\rangle + a|0\rangle)$. As discussed earlier, the loss is then modelled by a beam splitter, as shown in Fig.~\ref{fig:scheme}, with the resulting mixed state given by Eq.~(\ref{rho}).
We then apply the displacement operator $\mathcal{D}(-\beta) = e^{\beta^* \hat{a} - \beta \hat{a}^{\dagger}}$, which can be achieved by mixing the state with a large local oscillator at a highly transmittive beam splitter \citep{paris1996displacement}. This gives:
\begin{align}
\rho &= \mathcal{D}(-\beta)\rho_{_{UCS}}\mathcal{D}^{\dagger}(-\beta) \\
 &= \mathcal{N}_{u}^2 \left[ |\sigma\rangle\langle \sigma| + a^2|-\beta\rangle\langle -\beta|  \vphantom{e^{-\alpha^{2}_{\mu}(a)/2}}\right. \notag\\
  &\quad \left. + a e^{-\alpha^{2}_{\mu}(a)/2} \left( e^{i\theta}|\sigma \rangle\langle -\beta | + e^{-i\theta} |-\beta\rangle\langle \sigma| \right) \right], \notag
  \label{rho_D}
\end{align}
where $\theta=\alpha_{\eta}(a) \beta \sin{\phi}$ and $\sigma=\alpha_{\eta}(a)e^{i\phi}-\beta$. We then count the number of particles in the state $\rho$ and use a Bayesian scheme to infer the phase $\phi$ and the precision with which it can be measured $\delta\phi$ (we explain this scheme in detail in \citep{knott2014attaining}).

\begin{figure}[t]
\centering
\includegraphics[scale=0.47]{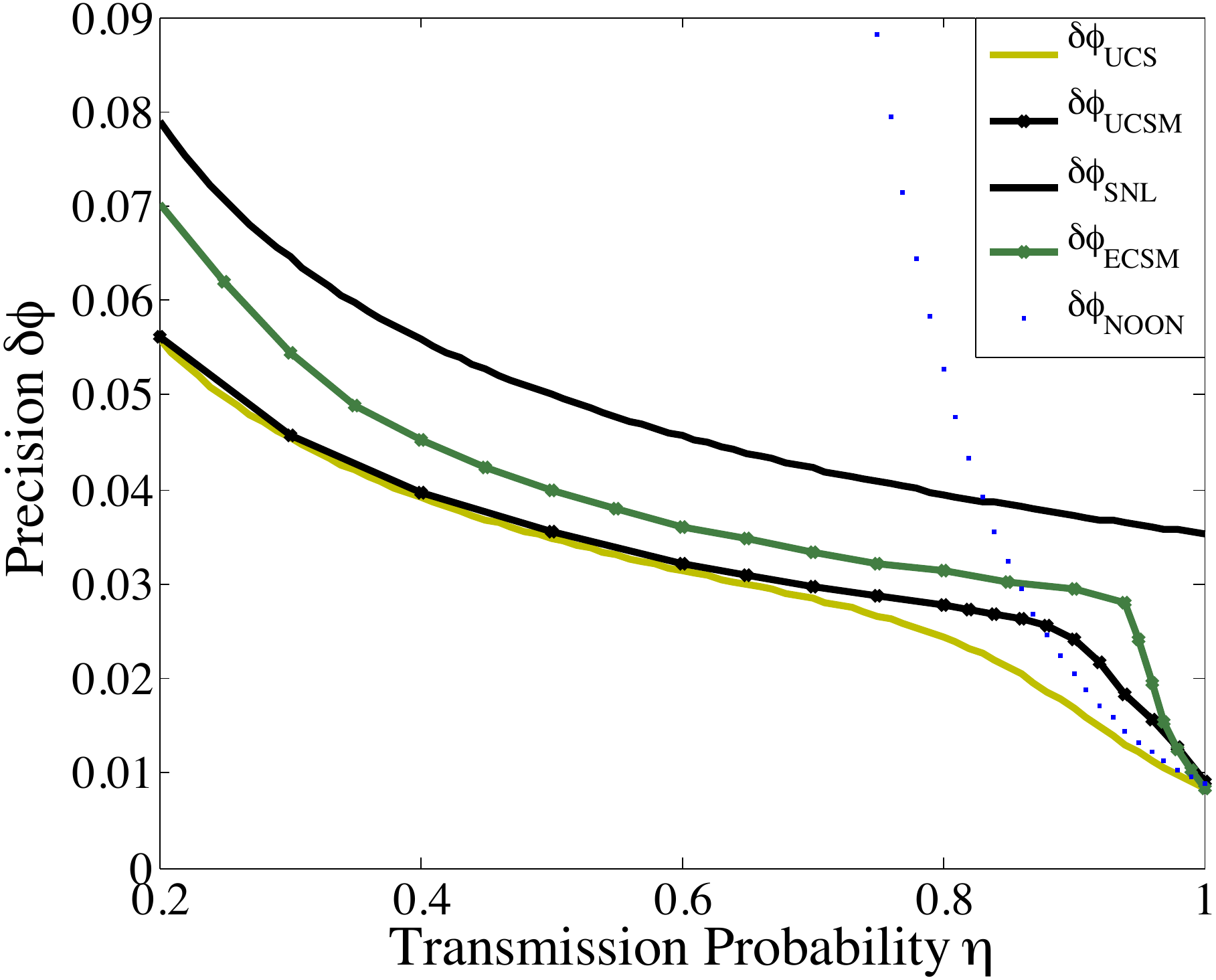}
\caption{(color online) Our measurement scheme, $\delta\phi_{\text{UCSM}}$, comes close the CRB for the UCS, $\delta\phi_{\text{UCS}}$, and shows large improvements over the ECS measurement scheme in \citep{knott2014attaining}, $\delta\phi_{\text{ECSM}}$. We see that our state surpasses the precision of the NOON state $\delta\phi_{\text{NOON}}$ and the SNL $\delta\phi_{\text{SNL}}$ for most loss rates. Here $\alpha_{bal}=4$.}
\label{fig:al4_graph_c}
\end{figure}

We obtain more precise measurements by taking $\beta > \alpha$, and we optimize over the phase $\phi$. The phase precision for this measurement scheme ($\delta\phi_{\text{UCSM}}$) is shown in Fig.~\ref{fig:al4_graph_c} (crossed black line) where $\beta = 4\alpha_{bal}$. Our scheme shows significant improvements over: the SNL; the ECS with the measurement scheme in \citep{knott2014attaining}; and the NOON state (for most loss rates). We see that our scheme is much more robust than the NOON state which is quickly destroyed when the transmission rate drops below $\eta = 0.9$. Whilst the CRB for the UCS shows only a small improvement over the (balanced) cat state, when we consider the measurement scheme the UCS is significantly better.

To understand why the UCS performs so well with this measurement scheme it is instructive to consider the case of a coherent state input, i.e $\ket{\Psi_{input}}=\ket{\alpha}$ in Fig~\ref{fig:scheme}. To find the phase precision for this input state and measurement we use the propagation of error formula:
\begin{eqnarray}
\delta\phi = \frac{\Delta\hat{X}}{|\frac{\partial \langle \hat{X} \rangle}{\partial \phi}|},
\end{eqnarray}
where $\Delta\hat{X} = \sqrt{ \langle \hat{X}^2 \rangle - \langle \hat{X} \rangle^2}$, and we take the number counting measurement operator $\hat{X}=\hat{a}^{\dagger}\hat{a}$. We find that the CRB, given by $\delta\phi_{\text{CS}}=1/\sqrt{2\alpha_{\eta}^2}$ with transmissivity $\eta$, is saturated in the limit $\beta \rightarrow \infty$, where $\beta$ is the displacement parameter. This is a $\sqrt{2}$ improvement over the generic scheme of a coherent state and a vacuum input fed into the arms of a standard Mach-Zehnder interferometer, which can measure at the SNL, $\delta\phi_{\text{SNL}}=1/\alpha_{\eta}$. We have found no explicit reference to this optimal measurement for a coherent state in the literature, but note that it bears a similarity to a heterodyne measurement, in which a large reference beam is used to amplify a signal to enhance precision. Since when $a=0$ a UCS reduces to a coherent state, $\delta\phi_{\text{CS}}$ is the upper bound on the phase precision that will be achieved with a UCS optimized over $a$. It is not clear how to get this close to saturating the bound for the balanced cat state, and so the UCS is significantly better when the measurement scheme is considered.

\emph{Conclusion -} High precision measurements are fundamental to physics, and quantum metrology offers a unique method for improving measurements beyond what is possible with classical physics. We show here that, for optical systems, multi-mode entanglement is not only unnecessary for phase estimation at the Heisenberg limit, it is actually detrimental to precision measurements when loss in included. Following this principle we introduce a single mode quantum superposition state: the unbalanced cat state. This state shows significant improvements over the alternatives, and can be created and precisely measured with present day, or near future, technology. We show that by tuning the degree of superposition in our state, and additionally by `chopping' our states into different sized chunks depending on loss rates, we can produce further improvements to our phase estimation scheme that allow us to surpass the precision obtained by multi-mode states. This work opens up a new approach to optical quantum metrology based on single-mode states which promises huge potential for future precision measurement protocols.\\

\begin{acknowledgments}
This work was partly supported by DSTL (contract number DSTLX1000063869). The authors work like to acknowledge recent work by Sahota and Quesada \cite{sahota2014quantum} who have shown independently that single mode squeezed states show advantages in optical quantum metrology in lossy systems.
\end{acknowledgments}

\bibliographystyle{apsrev}
\bibliography{MyLibrary_Cat2014}

\begin{thebibliography}{50}
\bibitem[{foo()}]{footnote1}
\bibinfo{note}{The SNL is often referred to as the \textit{scaling} $\delta\phi
  \propto 1/\sqrt{N}$, but we take the equality here as this is what is
  obtained in a classical interferometer set up.}
\end{thebibliography}

\end{document}